\documentclass[prb,superscriptaddress,twocolumn]{revtex4-2}
\usepackage{graphicx}
\usepackage{times,amsmath}
\usepackage{epsfig}
\usepackage{color}
\usepackage{graphicx}
\usepackage{dcolumn}
\usepackage{bm}
\usepackage{bookmark}
\usepackage{tabularx}
\usepackage{hyperref}
\usepackage{multirow}
\usepackage{array,mathtools,amssymb,booktabs,makecell}
\hypersetup{colorlinks=true, citecolor=blue, filecolor=blue, linkcolor=blue, urlcolor=blue}

\usepackage{float}
\usepackage{adjustbox}
\usepackage{upgreek}
\usepackage{soul}
\newcommand\titlelowercase[1]{\texorpdfstring{\lowercase{#1}}{#1}}

\makeatletter
\let\saved@includegraphics\includegraphics
\AtBeginDocument{\let\includegraphics\saved@includegraphics}
\renewenvironment*{figure}{\@float{figure}}{\end@float}
\makeatother

\begin{document}

\title{Topological singularity-induced Mott-like self-energy and its impact on Kondo cloud formation}




\author{Byungkyun Kang}
\email[]{bkang@udel.edu}
\affiliation{College of Arts and Sciences, University of Delaware, Newark, Delaware 19716, USA}


\author{Zachary Brown}
\affiliation{Department of Physics and Astronomy, Texas Tech University, Lubbock, Texas 79409, USA}


\author{Myoung-Hwan Kim}
\affiliation{Department of Physics and Astronomy, Texas Tech University, Lubbock, Texas 79409, USA}

\author{Hyunsoo Kim}
\affiliation{Department of Physics, Missouri University of Science and Technology, Rolla, 65409, MO, USA}

\author{Chul Hong Park}
\affiliation{Quantum Matter Core-Facility and Research Center of Dielectric and Advanced Matter Physics, Pusan National University, Busan 46240, Republic of Korea}

\begin{abstract}
We discovered that abnormal Mott physics can emerge even in weakly correlated 4$f$ fermions through their interplay with topological singularity.
Employing ab initio many-body perturbation theory combined with dynamical mean field theory, we show that 4$f$ electrons near the topological singular point experience strong effective Coulomb repulsion, as the hybridization channels are blocked near the singularity. As a result, Mott-like self-energy emerges, forbidding the coexistence of 4$f$ quasiparticles and the topological singularity at the same energy level in HoPtBi, PrPtBi, and PrAlGe.
The formation of 4$f$ quasiparticles is highly dependent on the energy of topological singularity relative to the Fermi level. 
This effect is suggested to be responsible for the selective quantum phenomena observed between heavy fermion behavior from Kondo resonance and anomalous transport from nontrivial topological states.


\end{abstract}

\maketitle

\section{I\titlelowercase{ntroduction}} 
Beyond topological insulators, topological semimetals are a class of materials that have gained significant attention due to the emergence of topological quasiparticles in the vicinity of Weyl points or quadratic band touchings. These materials exhibit intriguing and unconventional physical phenomena, making them promising candidates for the development of robust quantum devices~\cite{xu_prl2017,barry_science2016,lv_nphys2015}. Some of these phenomena include high Fermi velocities, Lorentz-violating type-II Weyl fermions, a strong anomalous Hall effect in ferromagnetic Weyl semimetals, and topological magnetic phases in topological semimetals~\cite{hu_apl2020,su_sciadv2017,kang_defect,yang_apl2019, yang_npjqm2022, daniel_ncom2020,pascal_prl2020}. Additionally, topological half-Heusler semimetals exhibit exotic physics such as tunable normal-state band inversion strength, superconducting pairing, magnetically ordered ground states, and anomalous Hall effects~\cite{yasuyuki_sciadv2015,chandra_pnas2018,jie_prb2021}.

The incorporation of strong electron correlation into topological systems is a burgeoning field of study that is expanding the frontiers of topological semimetals~\cite{robert_rpp2016}. The canonical example is topological Kondo insulators, where a topologically nontrivial state is developed through the interplay of strong correlations and spin-orbit interactions, as seen in SmB$_{6}$~\cite{maxim_prl2010,maxim_arcmp2016}. An experimental study suggests that the Kondo coherence of the flat $f$ bands has a dramatic influence on the low-temperature physical properties associated with the Berry curvature~\cite{byung_usbte}. A theoretical study has reported the breakdown of topological invariants in strongly correlated systems~\cite{jinchao_prl2023}. Recently, Weyl-Kondo semimetals, where the Kondo interaction renormalizes the bands hosting Weyl points, have been discovered~\cite{gau_ncom2018,sami_pnas2021,sami_ncom2022}. 
In a Kondo lattice, 4$f$ electrons that were localized can become quasiparticles, resulting in incoherent Kondo hybridization with conduction electrons at elevated temperatures~\cite{kang_ndnio2,byung_ute2,kang2023dual}. As the temperature decreases below the lattice coherence temperature ($T^{*}$), a Kondo insulator may appear due to the hybridization gap.
The hybridization between 4$f$ quasiparticles and conduction electrons leads to low carrier mobility, known as heavy fermions.
In the RPtBi (R = rare earth) half-Heusler family, emergent phenomena such as heavy fermion quantum criticality, unconventional superconductivity, and topologically nontrivial states are highly dependent on the choice of rare earth elements~\cite{eundeok_mrs2022}. The fundamental principles governing the interaction between topological quasiparticles and 4$f$ quasiparticles, which explains the diverse range of quantum phenomena, are still unknown.

\begin{figure}[ht]
\centering
\includegraphics[width=0.5
\textwidth]{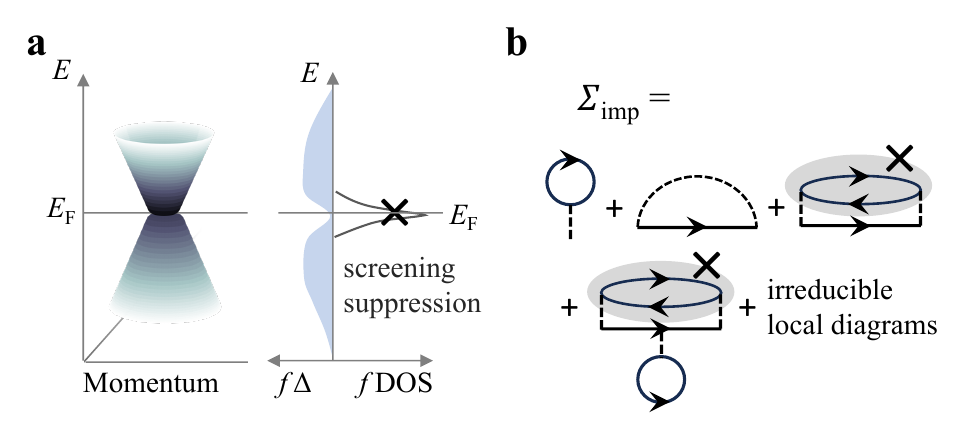}
\caption{\label{Fig_type}\
\textbf{Schematic diagram of the emergence of Mott-like self-energy.}
\textbf{a}, In 4$f$-based topological semimetals, the 4$f$ hybridization function ($\Delta$) primarily influences the screening of Coulomb repulsion between 4$f$ electrons, and is mainly affected by the density of states (DOS) of the topological bands. 
The topological singular point at the Fermi level with subtle DOS does not provide the necessary hybridization environment, leading to screening suppression and strong effective Coulomb repulsion. As a result, the formation of 4$f$ quasiparticles is hindered (indicated by a $\times$ symbol).
\textbf{b}, Diagrammatic representations of impurity self-energy within DMFT.
The solid lines denote the Green's function $G$, and the dashed lines represent interaction vertices, corresponding to the bare Coulomb interaction.
The "screening suppression" affects the self-energy, eliminating the particle-hole pair bubble ($\times$) in the diagrams.
Thus, the contribution of the first-order diagram, calculated with strong bare Coulomb repulsion to the self-energy is dominant, which should lead to a Mott-like impurity self-energy.
}
\end{figure}

The classification of strongly correlated quantum materials has been characterized by a diverse array of self-energy phenomena, stemming from a multitude of physical principles. These encompass Mott, Hund~\cite{byung_lanio2}, and Kondo~\cite{byung_ute2} physics, all of which are governed by the interplay of Coulomb and exchange interactions, as well as the localization of correlated electrons.
In this paper, we report the discovery of a topological singularity-induced Mott-like self-energy of 4$f$ electrons, which should be responsible for distinct quantum phenomena in rare-earth based topological semimetals. We revealed that topological singular points disrupt the essential hybridization channels of 4$f$ electrons, resulting in strong effective Coulomb repulsion $U_{\textrm{eff}}$ that closely resembles the bare Coulomb repulsion (see Fig.~\ref{Fig_type}). This effect eliminates the screening process of a particle-hole pair bubble in the Feynman diagrams of the self-energy (Fig.~\ref{Fig_type}b). Consequently, the self-energy takes on a Mott-like form with a strong on-site Coulomb repulsion, causing the 4$f$ electrons to be pushed away from the singularity.
The self-energy is centered at the energy level of topological singularity.
Thus, the topological semimetal HoPtBi does not have the Kondo resonance due to the Mott-like self-energy induced by topological singular points at the Fermi level ($E_\textrm{F}$). These semimetals exhibit anomalous transport that originates from nontrivial band topology. In PrPtBi, in comparison to LaPtBi, the topological singular points are positioned above $E_\textrm{F}$, allowing for Kondo hybridization between the 4$f$ quasiparticles and conduction electrons at $E_\textrm{F}$, leading to heavy fermion behavior.
We generalize this discovery by introducing PrAlGe and comparing it to LaAlGe in the End Matter.


\section{T\titlelowercase{opological singularity-induced} M\titlelowercase{ott-like self-energy}}
\begin{figure}[ht]
\centering
\includegraphics[width=0.5 \textwidth]{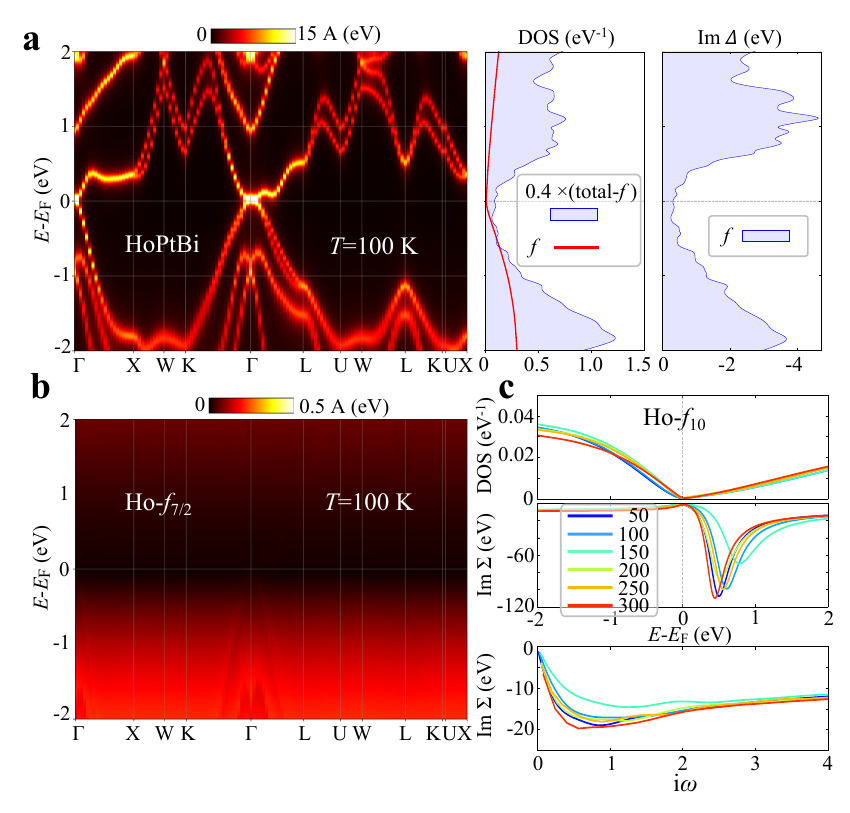}
\caption{\label{Fig_hoptbi}\
\textbf{Topological singularity-induced Mott-like self-energy in HoPtBi.} 
\textbf{a}, Spectral functions of HoPtBi at $T$=100 K (left panel), projected DOS of Ho-4$f$ and the total DOS with the Ho-4$f$ component subtracted (middle panel), and the imaginary part of the hybridization function ($\Delta$) of Ho-4$f$.
\textbf{b}, $f_{7/2}$ projected spectral functions at $T$=100 K.
\textbf{c}, $f_{10}$ projected DOS, imaginary part of self-energies on the real frequency axis and on the imaginary frequency axis, at varying temperature in unit of Kelvin.
}
\end{figure}

The linearized quasiparticle self-consistent GW + dynamical mean field theory (LQSGW+DMFT) method~\cite{choi2019comdmft} was utilized to calculate the electronic structure of rare-earth-based topological semimetals.
Simulations of HoPtBi revealed a distinct self-energy form, exhibiting a pronounced interplay between 4$f$ quasiparticles and topological singular point. Fig.~\ref{Fig_hoptbi}a shows a topological singular point at the $\Gamma$ point at $E_\textrm{F}$, while the Ho-4$f$ orbitals' projected spectral weight is prohibited at $E_\textrm{F}$, as shown in Fig.~\ref{Fig_hoptbi}b. Table~\ref{table_occ} and Fig. S3 demonstrate that the six $f$ orbitals in the $f_{5/2}$ multiplet are completely filled and exhibit similar behavior, while the eight $f$ orbitals in the $f_{7/2}$ multiplet are partially filled. Fig.~\ref{Fig_hoptbi}c illustrates that the DOS for Ho-$f_{10}$, which is representative of the $f_{7/2}$ multiplet, is zero at $E_\textrm{F}$. 

The calculated quasiparticle weight determined by the gradient of the imaginary component of the self-energy, $Z= 1/(1-\frac{\partial \text{Im}\Sigma(i\omega)}{\partial i\omega}\Bigr|_{i\omega\rightarrow0^{+}})$ is presented in Table~\ref{table_occ}. 
As shown in Fig.~\ref{Fig_hoptbi}c, the energy level of the topological singular point remains at $E_\textrm{F}$, and the magnitude of self-energy of Ho-4$f$ on the real axis remains strong upon cooling.
In general, within intermediate on-site Coulomb interaction ($U_{\textrm{C}}$), as temperature is lowered, Mott-like physics with strong self-energy diminishes, promoting quasiparticle formation at $E_\textrm{F}$~\cite{kang_ndnio2}. However, the self-energy of Ho-4$f$ on the real axis remains strong.
The calculated $Z$ factor remains at $\sim$0.03 in the temperature range, indicating that strong correlation persists, pushing the $f$ DOS away from $E_\textrm{F}$ and maintaining the topological singular point at $E_\textrm{F}$ at low temperature.
This indicates that the self-energy of Ho-4$f$ is still influenced by Mott-like physics. This effect is not solely caused by $U_{\textrm{C}}$, as evident from the modest static $U_{\textrm{C}}(\omega=0)$ of 2.03 eV for Ho-4$f$ (Fig. S2). The interplay between 4$f$ quasiparticle and the topological singular point is responsible for the majority of the Mott-like self-energy of Ho-4$f$ in HoPtBi.

\begin{figure}[ht]
\centering
\includegraphics[width=0.5
\textwidth]{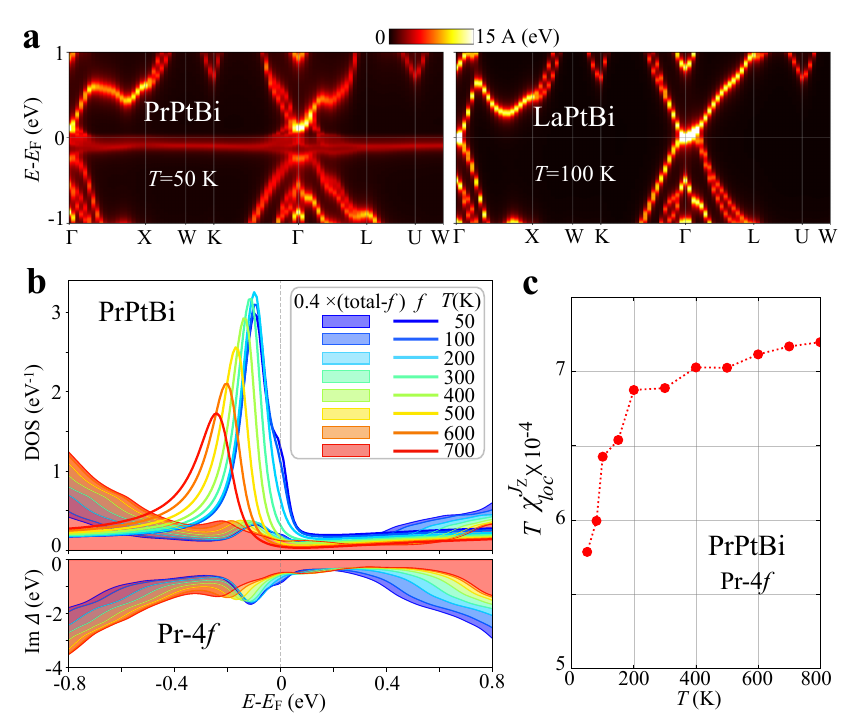}
\caption{\label{Fig_prptbi}\
\textbf{Effect of temperature on the electronic structure of PrPtBi.} 
\textbf{a}, Spectral functions of PrPtBi at 50 K and LaPtBi at 100 K.
\textbf{b}, Projected DOS of Pr-4$f$, the total DOS with the Pr-4$f$ component subtracted, and the imaginary part of the hybridization function ($\Delta$) of Pr-4$f$.
\textbf{c}, Product of temperature and local total angular momentum susceptibility $T\chi_{loc}^{J_Z}$ for Pr-4$f$.
}
\end{figure}

\begin{table*}[]
\caption{The electron occupation of Rare Earth-4$f$ orbitals and the $Z$ factor were calculated. The $Z$ factor was not presented for unoccupied orbitals. The $Z$ factor was presented only when it was greater than zero. The 4$f$ orbitals are labelled for convenience in this work.
The numbers within the brackets represent the Z factors, while the other numbers indicate the occupations of the orbitals.}\label{table_occ}
\scriptsize
\begin{center}
\scalebox{1}{
\begin{ruledtabular}
\vspace*{5mm}
\begin{tabular}{c|ccccccc|cccccccc}

 \multicolumn{1}{c|}{} &
 \multicolumn{15}{c}{4$f$ occupancy ($Z$ factor)} \\
 \hline
  $j$ & \multicolumn{7}{c|}{5/2} & \multicolumn{8}{c}{7/2}  \\
  \hline
  $j_{z}$&  &-2.5& -1.5 & -0.5 & 0.5 & 1.5 & 2.5 & -3.5 & -2.5 & -1.5 & -0.5 & 0.5 & 1.5 & 2.5 & 3.5 \\
  \hline
  label & &$f_{1}$& $f_{2}$ & $f_{3}$ & $f_{4}$& $f_{5}$ & $f_{6}$ & $f_{7}$ & $f_{8}$ & $f_{9}$ & $f_{10}$ & $f_{11}$ & $f_{12}$ & $f_{13}$ & $f_{14}$ \\

  \hline
  \multirow{4}{*}{HoPtBi} & T=100 K & 0.98 & 0.98 & 0.98 & 0.98 & 0.98 & 0.98 & 0.54 & 0.52& 0.53 & 0.50 & 0.51 & 0.47 & 0.48 & 0.47 \\
  &  & (0.51) & (0.55) & (0.59) & (0.58) & (0.55) & (0.56) & (0.04) & (0.03) & (0.03) & (0.02) & (0.02) & (0.01) & (0.01) & (0.01) \\
  & T=300 K & 0.97 & 0.96 & 0.96 & 0.97 & 0.97 & 0.97 & 0.52 & 0.52 & 0.51 & 0.48 & 0.56 & 0.51 & 0.51 & 0.51 \\
  &  & (0.36) & (0.67) & (0.41) & (0.38) & (0.32) & (0.43) & (0.03) & (0.03) & (0.03) & (0.02) & (0.04) & (0.03) & (0.03) & (0.03) \\

  \hline
  \multirow{4}{*}{PrPtBi} & $T$=100 K & 0.30 & 0.27 & 0.29 & 0.35 & 0.28 & 0.33 & 0.01 & 0.01 & 0.01 & 0.01 & 0.01 & 0.01 & 0.01 & 0.01 \\
  &  & (0.12) & (0.13) & (0.15) & (0.15) & (0.11) & (0.13) &  &  &  &  &  &  &  &  \\
  & $T$=300 K & 0.34 & 0.24 & 0.35 & 0.35 & 0.24 & 0.32 & 0.01 & 0.01 & 0.01 & 0.01 & 0.01 & 0.01 & 0.01 & 0.01 \\
  &  & (0.12) & (0.09) & (0.12) & (0.12) & (0.09) & (0.12) &  &  &  &  &  &  &  &  \\
  
  \hline
  \multirow{4}{*}{PrAlGe} & $T$=100 K & 0.43 & 0.42 & 0.01 & 0.01 & 0.42 & 0.44 & 0.01 & 0.01 & 0.01 & 0.00 & 0.00 & 0.01 & 0.01 & 0.01 \\
  &  & (0.11) & (0.13) & (0.13) & (0.13) & (0.13) & (0.12) &  &  &  &  &  &  &  &  \\
  & $T$=300 K & 0.38 & 0.37 & 0.16 & 0.16 & 0.36 & 0.38 & 0.01 & 0.01 & 0.01 & 0.01 & 0.01 & 0.01 & 0.01 & 0.01 \\
  &  & (0.12) & (0.13) &  &  & (0.13) & (0.12) &  &  &  &  &  &  &  &  \\
    
\end{tabular}
\end{ruledtabular}}
\end{center}
\end{table*}

\section{E\titlelowercase{nhanced Coulomb repulsion through screening suppression}}
Figure. S2 shows the dynamical $U_{\textrm{C}}$. In HoPtBi, a small static $U_{\textrm{C}}$ for Ho-4$f$ exerts a significant local screening effect on the local dynamics, ultimately enhancing the probability of 4$f$ quasiparticle formation. The pronounced frequency dependence of $U_{\textrm{C}}$ for Ho-4$f$ within the low-frequency range suggests the presence of a narrow and distinct 4$f$ quasiparticle peak at $E_\textrm{F}$~\cite{casula_prl2012}. However, as shown in Fig.~\ref{Fig_hoptbi}a, the DOS of neighboring orbitals in the vicinity of $E_\textrm{F}$ approaches zero as a result of the formation of the topological singular point. This leads to a substantial reduction in 4$f$ hybridization functions, that describe the extent to which 4$f$ orbitals are mixed with adjacent orbitals, and consequently, the shielding of the 4$f$ Coulomb repulsion~\cite{kristjan_prb2014}. Hence, the 4$f$ quasiparticle located at $E_\textrm{F}$ is exposed to negligible screening and undergoes a significant effective Coulomb interaction $U_{\textrm{eff}}$ (see Fig.~\ref{Fig_type}a) leading to a Mott-like phase transition.

As shown in Fig.~\ref{Fig_hoptbi}c, the self-energy decreases at temperatures from 300 to 150 K due to the weakening of Mottness.
However, the self-energy increases at temperatures from 150 to 50 K due to the enhanced screening suppression effect, which is accompanied by a higher probability of the formation of 4$f$ quasiparticles at lower temperatures.
As shown in Fig.~\ref{Fig_type}b, the proper sign of each term in the self-energy is obtained by multiplying it by the factor $(-1)^{h+1}$, where $h$ is the number of hole lines in the diagram. Therefore, the Fock term has a high magnitude with a negative sign. Higher-order diagrams can contribute positively or negatively to the self-energy, depending on the value of $h$ in the diagram. Starting with a large self-energy from the Fock term, the self-energy decreases as higher-order diagrams are taken into account. 
If higher-order terms are eliminated, the self-energy will maintain its large value.
Thus, the substantial Mott-like self-energy can emerge by the screening suppression. 
We use the term 'Mott-like' because Mott physics is reflected in the half-filling occupancy (Table I) and the emergence of Mott characteristics in both the upper and lower Hubbard incoherent bands (Fig. S3). Similar to the divergent self-energy at the Fermi level in a Mott insulator, the Mott-like self-energy appears as a divergent peak on the real frequency axis. Due to the energy level at which the interaction between the topological singularity and the 4$f$ electrons occurs being slightly above the Fermi level, the pole of the Mott-like self-energy on the imaginary frequency axis is located above $i\omega=0$.

\section{H\titlelowercase{eavy-fermion behavior in} P\titlelowercase{r}P\titlelowercase{t}B\titlelowercase{i}}
Unlike HoPtBi and LaPtBi, there are no experimental observations of a topological phase for PrPtBi~\cite{eundeok_mrs2022}, implying that the topological singularity may not be close to $E_\textrm{F}$.
As shown in Fig~\ref{Fig_prptbi}a, our calculations show that
in PrPtBi, the topological singular point
is located above $E_\textrm{F}$, while in LaPtBi, it is located at $E_\textrm{F}$. 
This energy level of topological singularity, which depends on the rare-earth element, may explain the inconsistent topological phases in RPtBi~\cite{eundeok_mrs2022}.
Here, we focus on the impact of the interplay between 4$f$ electrons and singularity on the formation of 4$f$ quasiparticles when the energy level of the singularity is above $E_\textrm{F}$ in PrPtBi.

Figure~\ref{Fig_prptbi}b shows the projected DOS of Pr-4$f$ (pDOS$_f$) and the total DOS subtracted by the pDOS$_f$ in PrPtBi. The minimum of the latter at the topological singular point coincides with the minimum of the 4$f$ hybridization function, indicating blocked hybridization channels at the topological singularity. This minimum shifts towards $E_\textrm{F}$ as the temperature decreases, and then remains around 0.15 eV.
Similar to the above-discussed HoPtBi, this minimum leads to a topological singularity-induced self-energy of 4$f$ electrons below 100 K, as shown in Fig. S4.
The energy level of the singular point is located slightly above $E_\textrm{F}$, and thus the interplay between the 4$f$ quasiparticles in the vicinity of $E_\textrm{F}$ and the topological singular point is relatively weak. This leads to weak self-energy and allows for the formation of a Pr-4$f$ peak at $E_\textrm{F}$.
As shown in Fig. S4, with a reduced $U_\textrm{C}$ at lower temperatures, there is a strong tendency for the formation of 4$f$ quasiparticles, which leads to a shift of the singularity below the Fermi level.
The flat Pr-4$f$ bands at $E_\textrm{F}$ can hybridize with the conduction electrons, resulting in the Kondo screening phenomenon~\cite{byung_ute2}. Table~\ref{table_occ} demonstrates a significant occupancy of Pr-4$f$ orbitals in the $j=5/2$ multiplet. The total occupation of Pr-4$f$ orbitals, calculated for both temperatures, is 1.9, implying that the substantial local magnetic moment of Pr-4$f$ is subjected to interaction by conduction electrons at low temperatures.

In order to explore the heavy-fermion behavior in PrPtBi, we computed the local total angular momentum susceptibility, which is expressed as
\begin{equation}
  \begin{split}
\chi_{\mathrm{loc}}^{J_Z}=\int_0^\beta d\tau\langle J_z(\tau)J_z(0)\rangle.
  \end{split}
\end{equation}
Fig.~\ref{Fig_prptbi}c shows the product of temperature and $\chi_{\mathrm{loc}}^{J_Z}$ of Pr-4$f$ as a function of temperature. A deviation from Curie-Weiss behavior is observed above 800 K, with a significant deviation at 200 K indicating the onset of the strong interaction with conduction electrons. The non-conformity of the spin susceptibility to the Curie law can be an indication of a heavy-fermion system~\cite{khodel_prl2005}. Our findings suggest that the heavy-fermion effect may be responsible for the non-metallic behavior of the electrical resistivity ($T < 300$ K) of PrPtBi~\cite{kasaya_phyb2000}.

\section{I\titlelowercase{ncompatibility between the 4$f$ quasiparticle and the topological singular point at} $E_\textrm{F}$.}

\begin{figure}[ht]
\centering
\includegraphics[width=0.4
\textwidth]{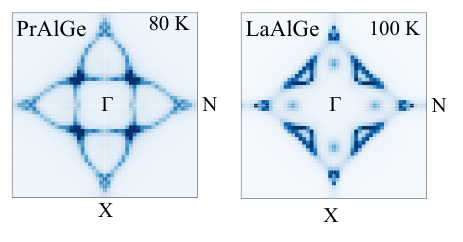}
\caption{\label{Fig_s_pralge_fs} \textbf{Fermi surface of PrAlGe and LaAlGe.} 
Fermi surfaces of LaAlGe and PrAlGe at $T$=100 and 80 K, respectively.}
\end{figure}

We have demonstrated a unique interplay between 4$f$ electrons and topological singular point arising from quadratic bands touching. To further investigate this interplay with other types of topological singular point, we have conducted a study on PrAlGe, a semimetal with a topological singular point originating from a linear band crossing near $E_\textrm{F}$. 
The spectral function of LaAlGe closely resembles the LQSGW bands, which exhibit a renormalization of the DFT bands (data not shown). This suggests that there is not a significant level of strong correlation in LaAlGe, as it does not contain 4$f$ electrons. Nevertheless, when the correlated Pr-4$f$ electrons in PrAlGe are treated within the DMFT approach, there is a modification to the band structure, particularly near the $\Gamma$ point, compared to the results obtained from DFT and LQSGW methods. As shown in Fig.~\ref{Fig_s_pralge_fs}, both the Fermi surfaces of LaAlGe and PrAlGe are consistent with those measured~\cite{su_sciadv2017,daniel_ncom2020}, and there is a significant contrast between the Fermi surface of LaAlGe and that of PrAlGe. These findings suggest that the Pr-4$f$ electrons are located near $E_\textrm{F}$, significantly impacting the electronic configuration of PrAlGe. 


\begin{figure}[ht]
\centering
\includegraphics[width=0.5 \textwidth]{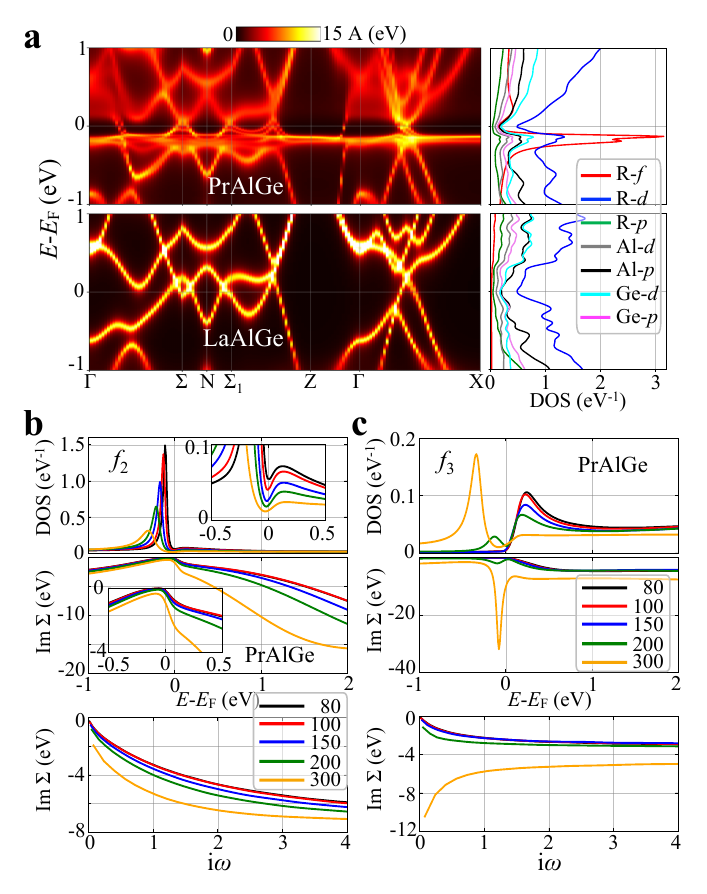}
\caption{\label{Fig_pralge}\
\textbf{Skipping over the Fermi level of 4$f$ electrons in PrAlGe.} 
\textbf{a}, Calculated spectral functions and orbital projected DOS of PrAlGe and LaAlGe at simulation temperatures of 100 K. The $p$ and $d$ orbital projected DOS were multiplied by 3 for visualization.
\textbf{b}, $f_2$ and \textbf{c}, $f_3$ projected DOS, imaginary part of self-energies on the real frequency axis, and on the imaginary frequency axis of PrAlGe. The temperature unit is Kelvin.
In (\textbf{b}), the inset shows magnified views of DOS and self-energy in the vicinity of the Fermi level.
}
\end{figure}

Figure~\ref{Fig_pralge}a shows a Weyl node at $E_\textrm{F}$ at the $\Sigma$ high-symmetry point, corresponding to the Weyl node W1 reported in previous studies~\cite{quoqiang_prb2018,su_sciadv2017}.
The Pr-4$f$ orbital-resolved DOS and self-energy are shown in Fig.~\ref{Fig_pralge}b and c. As the temperature decreases, the DOS for $f_2$ increases, and its peak position shifts towards $E_\textrm{F}$, manifesting a tendency to converge at $\sim$-0.2 eV. Of greater significance is the fact that the $f_2$ DOS is suppressed at $E_\textrm{F}$ across all temperature ranges. The quasiparticle peak is augmented as the temperature is lowered, due to a decrease in the self-energy, as demonstrated by the imaginary part of the self-energy. As illustrated in the inset of the imaginary self-energy on the real axis, similar to the $f_2$ DOS, the shoulder-like self-energy shows a tendency to converge below 150 K. This suggests that the emergence of Mott-like self-energy at lower temperatures, derived from $U_{\textrm{eff}}$, is responsible for pushing the 4$f$ quasiparticles away from the Weyl node.

Due to the Mott-like self-energy, a captivating phenomenon is observed in $f_3$, which bypasses $E_\textrm{F}$ by lowering the temperature. At 300 K, the imaginary part of the self-energy on the imaginary frequency axis exhibits a singularity, as shown in Fig.~\ref{Fig_pralge}c. This implies that the $f_3$ at high temperatures is governed by Mott physics, resulting in a divergent peak of the self-energy on the real frequency axis at $E_\textrm{F}$ and the formation of a pseudogap in the $f_3$-driven partial DOS. The $f_3$ exhibits behavior akin to that of Nd-4$f$ at elevated temperatures in NdNiO$_2$~\cite{kang_ndnio2}. Nevertheless, a distinguishing feature is that $f_3$ in PrAlG at 300 K exhibits the suppressed DOS in the vicinity of $E_\textrm{F}$. At 200 K, the self-energy decreases, resulting in the $f_3$ DOS peak moving towards $E_\textrm{F}$. The occupied DOS decreases while the unoccupied DOS increases, leaving the suppressed DOS at $E_\textrm{F}$. Below 150 K, $f_3$ remains unoccupied. The weight of $f_3$ is redistributed to other partially occupied $f$ orbitals, as shown in Table~\ref{table_occ}.

\section{C\titlelowercase{onclusion}}
We found that in rare-earth based topological semimetals, the formation of a 4$f$ quasiparticle and the shape of the 4$f$ DOS are highly dependent on the energy level of the topological singularity, indicating that the suppression of screening originates from the topological singularity.
In the case of HoPtBi, where the topological singularity is at the Fermi level, instead of the formation of a 4$f$ quasiparticle (which is favorable in weakly correlated systems with a static $U_\textrm{C}$ of 2.0 eV for Ho-4$f$), it exhibits a Mott-like DOS of Ho-4$f$ centered at the energy level of the topological singularity. The upper and lower Hubbard-like bands are split from the energy level of the singularity, indicating that a weak $U_\textrm{C}$ cannot lead to the Mott-like DOS of Ho-4$f$ without suppressing the screening by the singularity. In the case of PrAlGe, the 4$f$ electrons skip over the Fermi level due to the screening suppression caused by the topological singularity at the Fermi level.
Due to the Mott-like self-energy originating from the screening suppression, there is no Kondo hybridization between Ho-4$f$ quasiparticles and conduction electrons, and no sign of 4$f$ band inversion, suggesting the subset of 4$f$ electron being Mott-like insulator.
In this case, nontrivial topological bands play a dominant role in low-energy excitations with minimal interference from 4$f$ electrons.
When topological singular points are away from the Fermi level as in PrPtBi, the abnormal screening suppression in the topological singularity does not affect the quasiparticle at the Fermi level, and the 4$f$ heavy-fermion behavior is supported.
The 4$f$ electron at the Fermi level may undergo Kondo hybridization with conduction electrons. Moreover, when the temperature falls below the Kondo temperature, the coherent Kondo hybridization between 4$f$ and conduction electrons can induce a gap and band inversion, leading to the formation of a topological insulating system.
This physics should be useful in designing topological superconductors and quantum computing devices in strongly correlated systems.


\section*{Acknowledgments}  
We acknowledge the High Performance Computing Center (HPCC) at Texas Tech University for providing computational resources that have contributed to the research results reported within this paper.
M.K. and Z.B acknowledge the support from the National Science Foundation (Grant No. NSF DMR-2317013).
C.H. Park acknowledges the support by the National Research Foundation of Korea (NRF) grant (Grant No. NRF-2022R1A2C1005548) and the Ministry of Education (grant No. 2021R1A6C101A429).

\bigskip
\textbf{Competing Interests} The authors declare no competing interests.

\bigskip
\textbf{Data availability} The data that support the ﬁndings of this study are available from the corresponding
authors upon reasonable request.

\bigskip
\textbf{Author contributions} 
B.K. designed the project. B.K. performed the LQSGW+DMFT calculations and conducted the data analysis. 
All authors wrote the manuscript, discussed the results, and commented on the paper.
\bibliography{ref}

\end{document}